\documentclass[review,12pt]{elsarticle}
\usepackage{geometry}
\usepackage{amsmath}
\usepackage{amsfonts}
\usepackage{hyperref}
\usepackage{lineno}
\usepackage{caption}
\usepackage{booktabs}
\usepackage{xcolor}
\usepackage{soul}
\usepackage{tabularx}
\journal{arXiv}

\begin{document}

\begin{frontmatter}
\title{Unraveling impacts of polycrystalline microstructures on ionic conductivity of ceramic electrolytes by computational homogenization and machine learning}
\author[tud]{Xiang-Long Peng\corref{cor1}}
\ead{xianglong.peng@tu-darmstadt.de}
\author[tud]{Bai-Xiang Xu}
\cortext[cor1]{Corresponding author}
\address[tud]{Mechanics of Functional Materials Division, Institute of Materials Science, Technische Universität Darmstadt, 64287 Darmstadt, Germany}
\begin{abstract}
The ionic conductivity at the grain boundaries (GBs) in oxide ceramics is typically several orders of magnitude lower than that within the grain interior. This detrimental GB effect is the main bottleneck for designing high-performance ceramic electrolytes intended for use in solid-state Lithium-ion batteries, fuel cells, and electrolyzer cells. The macroscopic ionic conductivity in oxide ceramics is essentially governed by the underlying polycrystalline microstructures where GBs and grain morphology go hand in hand. This provides the possibility to enhance the ion conductivity by microstructure engineering. To this end, a thorough understanding of microstructure-property correlation is highly desirable. In this work, we investigate numerous polycrystalline microstructure samples with varying grain and grain boundary features. Their macroscopic ionic conductivities are numerically evaluated by the finite element homogenization method, whereby the GB resistance is explicitly regarded. The influence of different microstructural features on the effective ionic conductivity is systematically studied. The microstructure-property relationships are revealed. Additionally, a graph neural network-based machine learning model is constructed and trained. It can accurately predict the effective ionic conductivity for a given polycrystalline microstructure. This work provides crucial quantitative guidelines for optimizing the ionic conducting performance of oxide ceramics by tailoring microstructures.
\end{abstract}


\end{frontmatter}

\section{Introduction}
Polycrystalline ionic conducting oxide ceramics are commonly found in engineering \cite{kreuer2003proton,huang2001oxide,meng2019recent}. Depending on the mobile charge carrier therein, they can be categorized into different types, e.g., proton conducting \cite{kreuer2003proton,meng2019recent}, oxygen ionic conducting \cite{huang2001oxide}, and alkali-ionic (e.g., lithium, sodium, and potassium ions) conducting \cite{sasano2020grain} ones. They are promising candidates as electrolytes in solid fuel and electrolyzer cells \cite{kreuer2003proton} and all solid-state batteries (ASSBs). In these energy storage and conversion application scenarios, one of the key properties of concern is the effective (macroscopic) ionic conductivity. Polycrystalline microstructures consist of grains and grain boundaries (GBs). Experimental evidence (e.g., \cite{kreuer2003proton,ma2014atomic,bowman2017enhanced,wu2017origin,duan2020proton,shirpour2012space,hagy2023effects,huang2023effect,peng2024atomistic}) suggests that common oxide ceramics exhibit high intrinsic ionic conductivity in the grain bulk. However, the ionic conductivity at GBs is typically orders of magnitude lower than the bulk one, resulting in unacceptable low effective or total conductivity. For detailed information regarding the magnitudes of GB, bulk, and total conductivity of common oxide ceramics, one is referred to the above mentioned references.

The low GB ionic conductivity are attributed to different factors, e.g. the GB compositions and structures \cite{sasano2020grain,bowman2017enhanced} and space charge layers \cite{de2011defect}. Numerous efforts have been devoted to improving the intrinsic GB conductivity \cite{kreuer2003proton,sasano2020grain,bowman2017enhanced,wu2017origin,sun2011lowering}. However, this challenging task remains unsolved. On the other hand, the effective properties of a microstructured material depend not only on the composition properties but also on the microstructural features. Thus, apart from the bulk and GB conductivities, the microstructural features relevant to grains and GBs, e.g., grain size, grain orientation, and grain and GB arrangement also strongly affect the effective ionic conductivity \cite{wu2017origin,yamazaki2009high,bowman2019linking}. Therefore, a thorough understanding of polycrystalline microstructure-effective conductivity correlation is crucial for designing oxide ceramics with enhanced effective ionic conductivity.

In addition to experiment investigations, analytical modeling and numerical simulations (e.g., \cite{hu2017phase, liu2021impedance}) are efficient tools for studying the ionic conducting behavior in oxide ceramics. The brick layer model \cite{verkerk1982effect} is one of the most widely used analytical models, which provides an explicit expression of the total conductivity as a function of the grain and GB conductivities, grain size, and GB thickness \cite{avila2010protonic,wang2014schottky}. It can be used in combination with impedance spectroscopy measurements to evaluate the GB conductivity based on the measured total conductivity \cite{wu2017origin}. Although it is easy to use, this idealized model ignores some important microstructural characteristics such as the inhomogeneous distributions of grain shape, grain size, and GB conductivities \cite{fleig1999impedance}. To account for these microstructure influences, numerical simulations are desired. Similar to the assumption in the brick layer model, one can consider that each grain and GB possess a constant ionic conductivity. Then, a steady-state conducting problem governed by the conservation of charge and Ohm's law in a realistic polycrystalline microstructure is to be solved numerically \cite{bielefeld2020modeling}. In this case, modeling of GB ionic conduction requires special attentions.

In recent years, data-driven or machine learning (ML) methods have been widely applied for surrogate modeling of microstructure-property correlations in microstructured materials including polycrystalline materials. Once trained and validated, data-driven models can efficiently predict the effective properties of a given microstructure. Thus, they can accelerate the process of designing microstructures with optimized properties. There are various ML models. Among them, the convolutional neural networks (CNNs) \cite{kondo2017microstructure,liu2022deep} and graph neural networks (GNNs)\cite{dai2021graph,dai2023graph,hestroffer2023graph} are commonly used to model polycrystalline microstructures. CNN models directly take artificial and/or experimental microstructure images or image sequences as inputs, preserving all microstructure details. However, due to the high dimensionality of these inputs, particularly for three-dimensional microstructures, CNN models are computationally expensive. On the other hand, the inputs of GNN models are grain and grain boundary (GB) feature matrices along with an adjacency matrix, which have significantly lower dimensions compared to image-based descriptions. As a result, GNN models are computationally more efficient. Furthermore, thanks to their innovative architecture, GNN models effectively capture inherent grain-grain and grain-GB interactions in polycrystalline microstructures. In fact, recent findings \cite{hestroffer2023graph} suggest that GNN models outperform other ML models both in terms of accuracy and efficiency when it comes to modeling polycrystalline microstructures.

Based on the above backgrounds, the contributions of this work are twofold: revealing the quantitative correlations between polycrystalline microstructure and effective ion conductivity in oxide ceramics and encapsulating these correlations into a GNN-based ML model. To this end, we propose a numerical framework to investigate the effective ionic conductivity of polycrystalline oxide ceramics. We conduct numerous simulations on polycrystals with varying grain and GB features to systematically investigate their influence on the effective conductivity. Based on the numerous data generated by simulations, we construct and train a GNN-based surrogate model to correlate the polycrystalline microstructure to its effective ion conductivity. These together provide quantitative guidelines on how to enhance the effective conductivity in ionic conducting ceramics by microstructure engineering.

\section{Model and methods}
\subsection{Polycrystalline microstructures: generation and characterization}
\begin{figure}[!htbp]
	\centerline{\includegraphics[width=\textwidth]{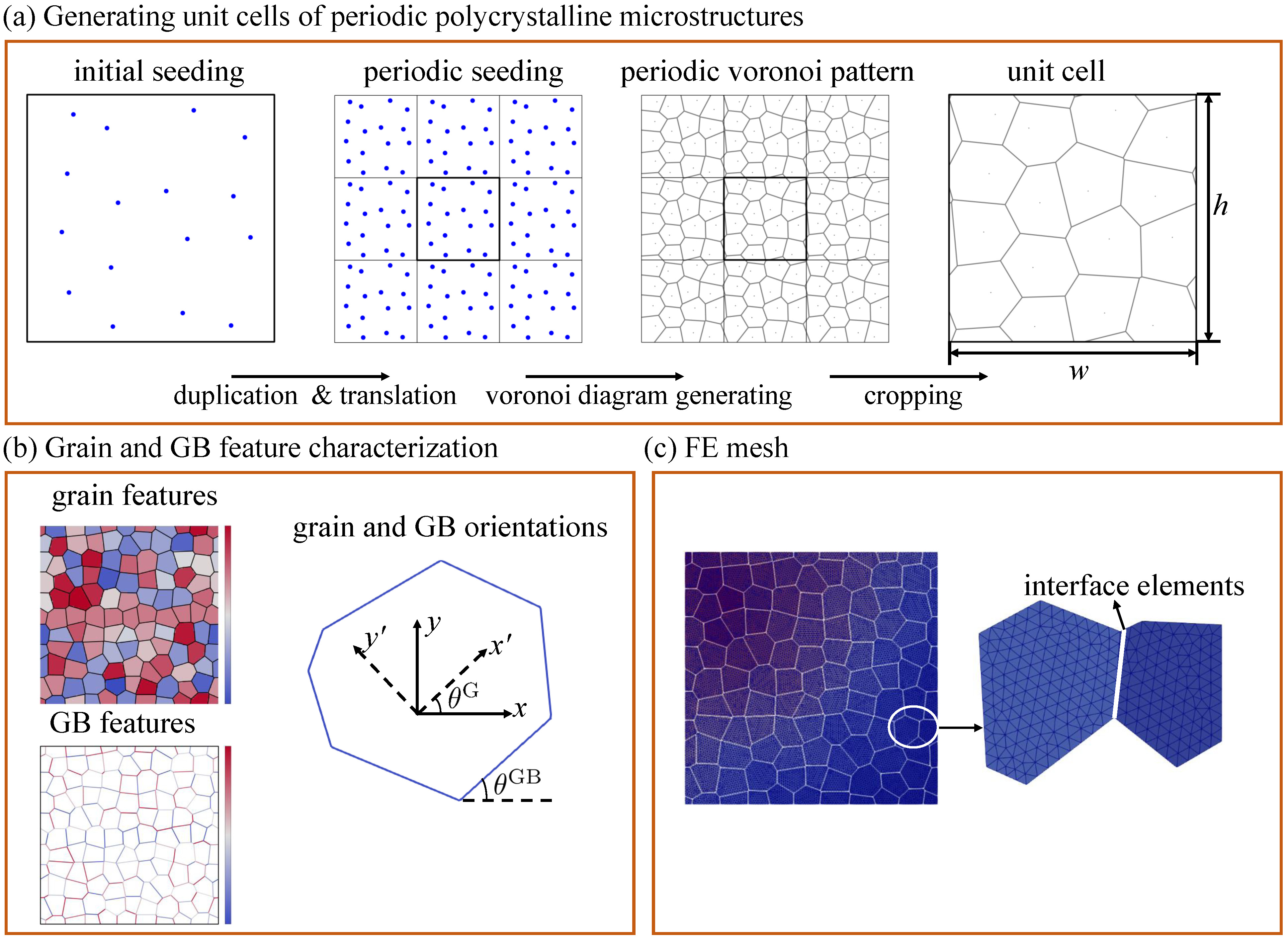}}
	\caption{Polycrystalline microstructure generation, characterization, and meshing. (a) The 2D Voronoi diagrams exploited to represent the 2D  polycrystalline microstructures. The microstructure generation processes include initial seeding, duplicating and translation of the initial domain to achieve periodic seeding, Voronoi tessellation, and cropping the final unit cell. (b) Schematic of grain and GB feature distributions. The grain and GB features are quantified for each microstructure, the influence of which on the effective ionic conductivity will be investigated. These microstructure features act as the inputs of the GNN-based surrogate model. (c) FE mesh of a unit cell for FE simulations. The grains and GBs are meshed with triangular elements and the two-node interface elements, respectively.}
	\label{geometry}
\end{figure}

As commonly adopted in the literature \cite{bargmann2018generation}, we use two-dimensional (2D) Voronoi diagrams to represent the 2D polycrystalline microstructures. As illustrated in Fig.\ \ref{geometry}a, the microstructure generation starts by randomly placing seeds in a 2D rectangular domain. The number of seeds is identical to the number of grains. To achieve periodic microstructures, the seeded domain is duplicated 8 times and translated to create a large rectangular domain, on which the Voronoi tessellation is performed. Finally, a unit cell is cropped which acts as the representative volume element (RVE) for the periodic polycrystalline microstructure. In the unit cell, the polygons denote grains and their edges are the GBs. The geometrical features of the microstructure such as the grain size and GB length are determined by the number and distribution of the seeds. As illustrated by \cite{zhu2001effects}, the degree of regularity of the generated polycrystals depends on the minimum distance $\delta$ between any two seeds. A larger $\delta$ corresponds to a more regular microstructure with more uniformly distributed grain sizes. In such a manner, polycrystals with different degrees of regularity are generated by choosing different $\delta$.

Given the geometry of a polycrystal RVE, the crystallographic parameters and material properties are assigned to each grain and GB randomly or in a controlled manner (see Fig.\ \ref{geometry}b). Numerous RVEs with different microstructural features are generated to study the influence of microstructures on the effective ionic conductivity. The above microstructure generation processes are implemented in the commercial software Matlab R2023a. 
\subsection{Ionic conduction model for grain interior}
\label{bulk_model}
The ionic conducting behavior in the grain bulk is modeled by the following Ohm's law \cite{hu2017phase,bielefeld2020modeling}, 
\begin{equation}
	j^{\text{m}}_{k}=\sigma_{kl}^{\text{G}}(\mathbf{x})E^{\text{m}}_{l}
	\label{bulk_Ohm}
\end{equation}
where $j^{\text{m}}_{k}$, $\sigma_{kl}^{\text{G}}$, and $E^{\text{m}}_{l}$ denote the current density vector, the bulk conductivity tensor, and the electric field vector, respectively. The electric field is defined as the gradient of the electric potential $\varphi^{\text{m}}$, i.e., $	E^{\text{m}}_{k}=-\partial\varphi^{\text{m}}/\partial x_{k}$ with $x_k$ being the coordinates. Here and hereafter, the superscripts 'm' and 'M' are used to denote quantities defined at the microscale (i.e., the microstructure level) and macroscale, respectively. According to the Nernst-Einstein equation \cite{nernst1888kinetik}, the ionic conductivity depends on the local ion diffusivity and concentration. In this work, we treat the bulk and GB ionic conductivities as known constants. 
In the steady state, the conservation of charge results in the following governing equation in the bulk,
\begin{equation}
	\frac{\partial j^\text{m}_{k}}{\partial x_k}=0.
	\label{bulk_govern}
\end{equation}

\subsection{Ionic conduction model for GB}
\label{GB_model}
We propose an interface model to characterize the ionic conducting at the GB. As the GB thickness is usually much smaller than the grain size, the distribution of the electric potential across the GB is assumed to be linear such that the current density $j^{\text{GB}}$ therein is proportional to the electric potential jump across the GB, i.e.,
\begin{equation}
	j^{\text{GB}}=-\frac{\sigma^\text{GB}}{t^\text{GB}}(\varphi^{\text{m}+}-\varphi^{\text{m}-}),
	\label{GB_ohm}
\end{equation}
where $\sigma^\text{GB}$ and $t^\text{GB}$ are the GB conductivity and thickness, respectively, and $\varphi^{\text{m}+}$ and $\varphi^{\text{m}-}$ are the electric potentials at both sides of the GB. $t^\text{GB}$ is considered to be the same for all GBs in a certain polycrystal. $\sigma^\text{GB}$ may differ among GBs \cite{bowman2019linking}. The conservation of charge in the steady state yields the following boundary conditions at the GB 
\begin{equation}
	j^{\text{m}-}_{k}n^{-}_{k}=	j^{\text{GB}},\quad
	j^{\text{m}+}_{k}n^{+}_{k}=	-j^{\text{GB}},
	\label{GB_govern}
\end{equation}
with $n^{-}$ and $n^{+}$ being the GB normals. The ionic conducting along the GB is not considered since the GB conductivity is much lower than the bulk one \cite{bowman2017enhanced,mcnealy2014use}.

The GB conductivity may depend on the GB characters such as the misorientation angle, the type, and the local chemical compositions \cite{bowman2019linking}. Here, we treat the GB conductivity as a known material parameter and focus on its influence on the effective conductivity.

\subsection{Orientation-dependence of ionic conductivity}
In general, the bulk ionic conductivity is anisotropic and hence orientation-dependent. For orthotropic crystals, only the diagonal components in the bulk conductivity tensor $\sigma_{kl}^{\text{G}}$ are non-zero in the crystallographic coordinate system. Thus, in the 2D case, the bulk conductivity is quantified by the two principal values $\sigma_{1}^{\text{G}}$ and $\sigma_{2}^{\text{G}}$. Then, the bulk conductivity in an individual grain with an orientation angle $\theta^\text{G}$ (see Fig.\,\ref{geometry}b) in the global coordinate system is expressed as
\begin{equation}
	\begin{bmatrix}
		\sigma^\text{G}_{11}
		& \sigma^\text{G}_{12}\\
		\sigma^\text{G}_{12}
		& \sigma^\text{G}_{22}\\
	\end{bmatrix}
	=	\begin{bmatrix}
		\text{cos}\theta^\text{G}& -\text{sin}\theta^\text{G}\\
		\text{sin}\theta^\text{G}& \text{cos}\theta^\text{G}
	\end{bmatrix}
	\begin{bmatrix}
		\sigma^\text{G}_{1}
		& 0\\
		0
		& \sigma^\text{G}_{2}\\
	\end{bmatrix}
	\begin{bmatrix}
		\text{cos}\theta^\text{G}& \text{sin}\theta^\text{G}\\
		-\text{sin}\theta^\text{G}& \text{cos}\theta^\text{G}
	\end{bmatrix}.
	\label{transform}
\end{equation}
In this work, we consider that $\sigma_{1}^\text{G}$ and $\sigma_{2}^\text{G}$ are the same for all grains in a polycrystal and the grain orientation angle $\theta^\text{G}$ varies among grains. 

\subsection{Effective ionic conductivity: computational homogenization}
\label{com_hom}
We focus on evaluating the effective or macroscopic ionic conductivity of a polycrystalline microstructure with given bulk and GB conductivities.
To this end, the computational homogenization method is exploited. We assume that the scale-separation holds such that the effective ionic conductivity can be calculated by performing simulations on an RVE. At the microscale, the ionic conducting behavior is characterized by the bulk and GB models in Sections~\ref{bulk_model} and \ref{GB_model}. At the macroscale, the polycrystalline material is effectively treated as a homogeneous medium. Its effective conducting behavior is modeled by the averaged Ohm's law, i.e.,
\begin{equation}
	\begin{split}
		j^{\text{M}}_{k}=\sigma^\star_{kl}E^{\text{M}}_{l}, 
	\end{split}
\label{eff_Ohm}
\end{equation}
where the macroscopic current density $j^{\text{M}}_{k}$ and electric field $E^{\text{M}}_{l}$ are defined as the volume-average of their counterparts at the microscale, i.e.,
\begin{equation}
j^{\text{M}}_{k}=\frac{1}{V}\int_{V}j^{\text{m}}_{k}\text{d}V, \quad
E^{\text{M}}_{k}=\frac{1}{V}\int_{V}E^{\text{m}}_{k}\text{d}V.
\end{equation}
To satisfy the Hill-Mandel condition $j^{\text{M}}_{k}E^{\text{M}}_{k}=\int_{V}j^{\text{m}}_{k}E^{\text{m}}_{k}\text{d}V/V$, the RVE is subjected to the following periodic boundary condition
\begin{equation}
		\varphi^{\text{m}(2)}-\varphi^{\text{m}(1)}=-E^{\text{M}}_k[x_k^{(2)}-x_k^{(1)}],
		\label{pbc}
\end{equation}
where the superscripts (1) and (2) denote the quantities at two periodic boundary points $x_k^{(1)}$ and $x_k^{(2)}$, respectively.

In the 2D case, the effective Ohm's law \eqref{eff_Ohm} can be rewritten in the matrix form as
\begin{equation}
	\begin{bmatrix}
		j^{\text{M}}_{1}
		\\
		j^{\text{M}}_{2}
	\end{bmatrix}
	=\begin{bmatrix}
		\sigma^{\star}_{11}
		& \sigma^{\star}_{12}\\
		\sigma^{\star}_{12}
		& \sigma^{\star}_{22}\\
	\end{bmatrix}
	\begin{bmatrix}
E^{\text{M}}_{1}
	\\
E^{\text{M}}_{2}
\end{bmatrix}.
\label{eff_Ohm2}
\end{equation}
Thus, to determine the three independent effective conductivity constants $\sigma^{\star}_{11}$, $\sigma^{\star}_{22}$ , and $\sigma^{\star}_{12}$, one needs to consider two loading cases. In each simulation, $E^{\text{M}}_{1}$ and $E^{\text{M}}_{2}$ are prescribed by the periodic boundary condition \eqref{pbc} to determine the distribution of $j^{\text{m}}_{1}$ and $j^{\text{m}}_{2}$, the volume average of which are $j^{\text{M}}_{1}$ and $j^{\text{M}}_{2}$.  Subsequently, the $\sigma^{\star}_{11}$, $\sigma^{\star}_{22}$ , and $\sigma^{\star}_{12}$ are evaluated by \eqref{eff_Ohm2}.

The numerical implementation of the above formulation is conducted by the finite element (FE) method. As illustrated in Fig.\ \ref{geometry}c, the grains and GBs are discretized into triangular elements and two-node interface elements\footnote{The element size is about 1/10 of the average grain size, which is sufficient to ensure mesh convergence.}, respectively.  Periodic mesh is generated in the open-source software Gmsh \cite{geuzaine2009gmsh}. Subsequently, the FE simulations are conducted in the open-source FE software MOOSE \cite{permann2020moose}. The interface elements are added by the built-in mesh generator in MOOSE. Python scripts are written to automatize the simulation processes, which enables efficiently simulating numerous microstructures.

The parameters involved in the simulations are the two principal bulk conductivities $\sigma_{1}^\text{G}$ and $\sigma_{2}^\text{G}$, the grain orientation angle $\theta$, the GB conductivity $\sigma^\text{GB}$, and the GB thickness $t^\text{GB}$.

The proposed bulk and GB models for ionic conduction, along with their numerical implementation, enable not only the determination of effective conductivity along an arbitrary direction but also the characterization of microscopic electric fields and current densities for a realistic polycrystalline microstructure. And the influence of grain and GB features on the effective conductivity are well captured. These aspects can not be adequately considered by the brick-layer model \cite{verkerk1982effect} and other existing numerical methods, e.g, \cite{fleig1999impedance,dezanneau2006grain,dezanneau2006effect,fleig2002grain}.

In the supplementary material, the verification of the numerical implementation of the proposed model by a benchmark problem is presented.

\subsection{The GNN-based ML model}
\label{surrogate model}
\begin{figure}[!htbp]
	\centerline{\includegraphics[width=\textwidth]{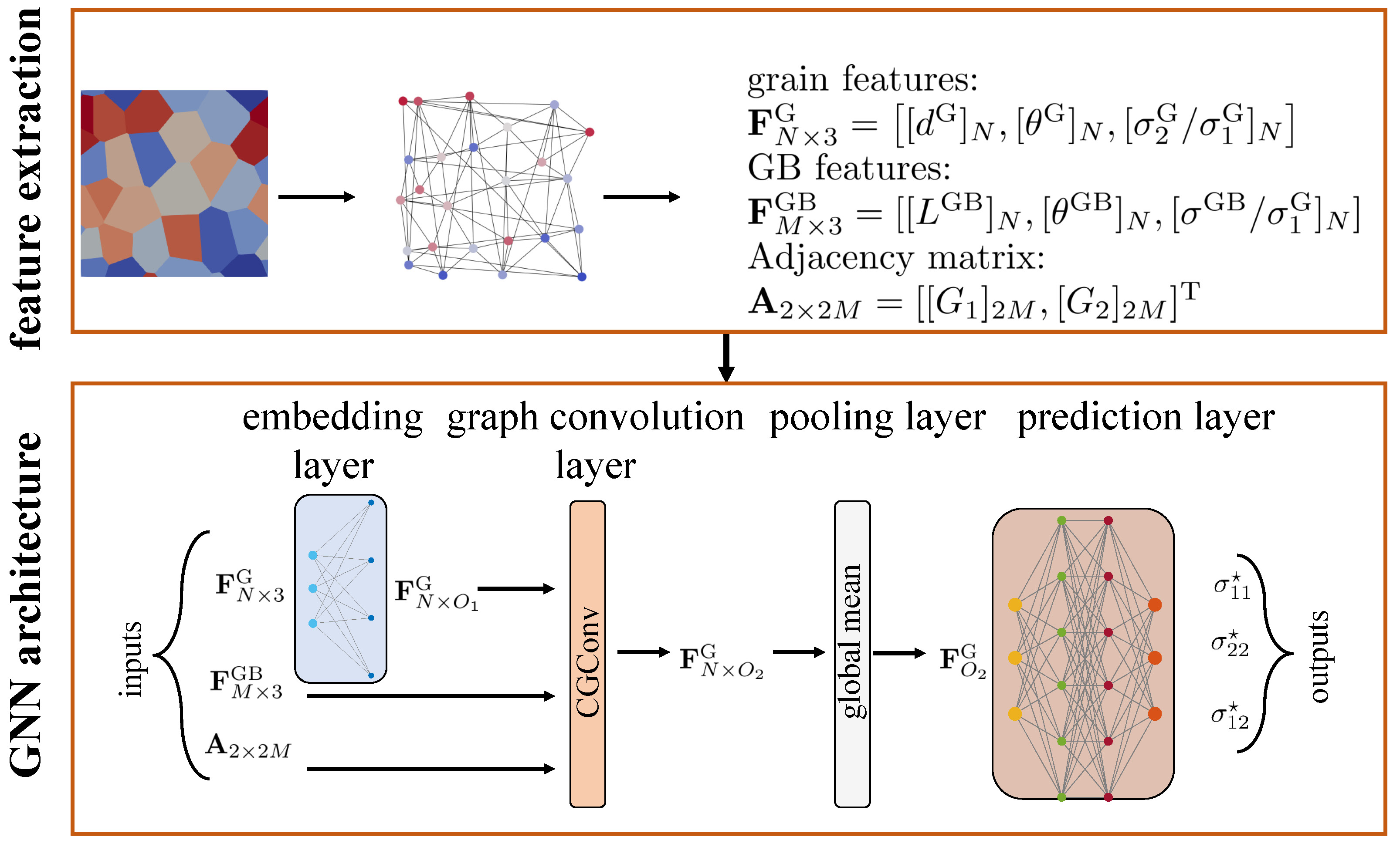}}
	\caption{GNN model architecture. (a) Converting a polycrystal RVE into a graph consisting of nodes and edges. Each grain (i.e., node) and each GB (i.e., edge) are characterized by three features. The grain and GB feature matrices together with the adjacency matrix act as the inputs to the GNN model. (b) The architecture of the GNN model. It consists of an embedding layer, a few convolutional layers, a pooling layer, and a prediction layer. For a given polycrystalline microstructure, the GNN model predicts the corresponding ionic conductivities.}
	\label{GNN}
\end{figure}
We construct a GNN-based ML model to predict the effective ionic conductivity for a given polycrystalline microstructure. As illustrated in Fig.\ \ref{GNN}, a polycrystal is first converted to a graph consisting of nodes (i.e., grains) connected by edges (i.e., GBs). Based on the graph, the grain (node) feature matrix, the GB (edge) feature matrix, and the adjacency matrix characterizing the grain connectivity are extracted, which act as the inputs to the GNN model. For each grain, we select three features, i.e., the normalized grain size $d^\text{G}/t^\text{GB}$, the grain orientation angle $\theta^\text{G}$, and the ratio between the two principal ionic conductivities $\sigma^\text{G}_{2}/\sigma^\text{G}_{1}$. Here, the grain size $d^\text{G}$ is defined as the square root of the grain area. For each GB, the three features are the normalized length $L^\text{GB}/t^\text{GB}$, the inclined angle $\theta^\text{GB}$ (see Fig.\,\ref{geometry}b), and the normalized conductivity $\sigma^\text{GB}/\sigma^\text{G}_{1}$. Thus, for a polycrystal with a total grain number $N$ and a total GB number $M$, the grain feature matrix, the GB feature matrix, and the adjacency matrix are $N\times 3$, $M\times 3$, and $2\times 2M$ matrices, respectively.

The first part of the GNN model is an embedding layer, which operates on each row (i.e., three features of each grain) of the grain feature matrix. Subsequently, the updated grain feature matrix together with the GB feature matrix and the adjacency matrix is forwarded to the convolutional layer, which is the core of a GNN model. In the convolutional layer, the grain features of each grain are further updated by combining features from all grains and GB connected to it. In such a manner, by introducing more than one convolutional layer, the interaction among grains and GBs that are not directly connected can be incorporated as well. Here, we choose the CGConv model proposed in \cite{xie2018crystal}, in which the convolution function is expressed as
\begin{equation}
\mathbf{x}_{i}^{\prime}=\mathbf{x}_{i}+\sum_{ j \in \mathcal{N}(i)}\sigma\left(\mathbf{z}_{ i ,j}\mathbf{W}_{ f }+\mathbf{b}_{ f }\right)\odot g \left(\mathbf{ z }_{ i,j}\mathbf{W}_{s} + \mathbf{b}_{s} \right)
\end{equation}
where $\mathbf{x}_{i}^{\prime}$ and $\mathbf{x}_{i}$ are the old and updated feature vectors of the $i$th grain, $\mathcal{N}(i)$ denotes the number of its total adjacent grains. $\mathbf{z}_{i,j}=\mathbf{x}_{i}\oplus\mathbf{x}_{j}\oplus\mathbf{e}_{i,j}$ is the concatenation of the two adjacent grain feature vectors and the feature vector $\mathbf{e}_{i,j}$ of the GB between them. $\sigma$ and $g$ are the activation function, and $\mathbf{W}_{i}$ and $\mathbf{b}_{i}$ are trainable weights and bias. The convolutional layer outputs the updated grain feature matrix, which is then compressed into a low-dimensional vector by the pooling layer. Here, the global mean pooling is used. Subsequently, the low-dimensional vector is forwarded to the property prediction layer which is essentially a feed-forward neural network. Finally, the outputs of the GNN model are the three effective ionic conductivities. 

The GNN model is constructed and trained by the deep learning platform Pytorch \cite{paszke2019pytorch}. All the relevant model parameters are summarized in Table 1. 
\begin{table}[htbp]
	\caption{List of parameters for GNN-based surrogate modeling. The GNN model parameters and the training-relevant parameters are summarized. }
	{\small
		\begin{tabular}{p{0.4\textwidth}p{0.6\textwidth}}
			\toprule
			GNN model architecture and \\ hyperparameters & Options\\					 
			\midrule
			Embedding layer & A single perceptron layer (3, 10; activation function: relu)\\
			Graph convolutional layer & 2 CGConv layers (activation function:  relu)\\
			Prediction layer & A feed-forward neural network with 2 hidden layers \newline (10, 128, 128, 3; activation function:  relu ) \\
			Optimization algorithm& Adam\\
			Learning rate& 0.002\\
			Epoch number & 400 \\
			Batch size & 40\\	
			\bottomrule
	\end{tabular}}
	\label{NN_p}
\end{table}
\section{Results and discussion}
Here, we investigate the influence of grain size, grain orientation, and the bulk and GB conductivities on the effective conductivity of polycrystals.
In addition, the training and validation of a GNN-based surrogate model for predicting the effective ionic conductivity of a given polycrystal is presented.

In the following, unless otherwise stated, $\sigma^\text{G}_{1}=\sigma^\text{G}_{2}=\sigma^\text{G}$ and $\sigma^\star=\sigma^\star_{11}$. The total grain number in each RVE is $n^\text{G}=100$. The height and width of the RVE are $w=h=\sqrt{n^\text{G}}\bar{d}^\text{G}$ with $\bar{d}^\text{G}$ being the nominal average grain size.
\subsection{Grain size-dependence of effective ionic conductivity}
\begin{figure}[!htbp]
	\centerline{\includegraphics[width=\textwidth]{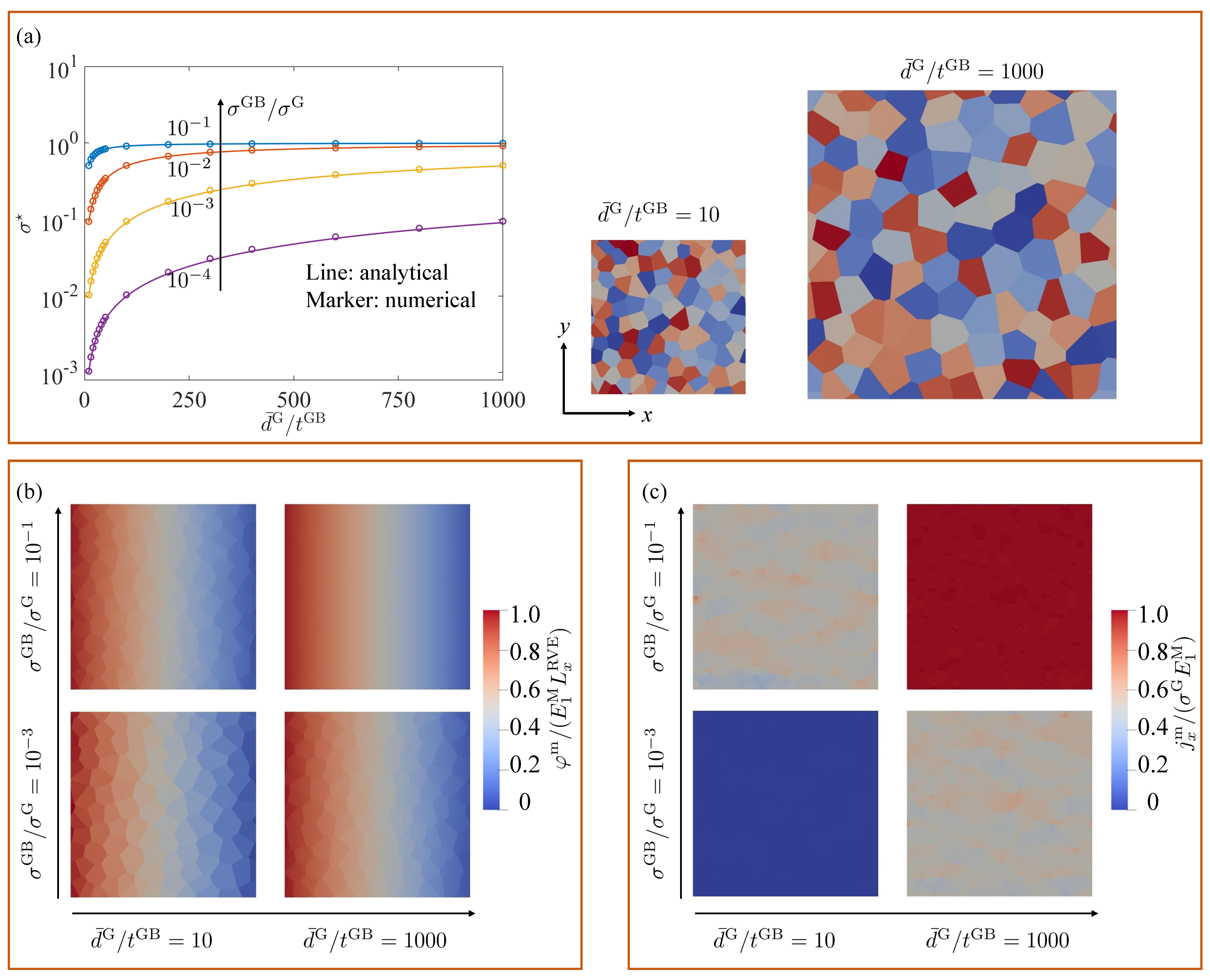}}
	\caption{Size-dependent effective conductivity. (a) The normalized effective conductivity $\sigma^\star/\sigma^\text{G}$ varying with the normalized grain size $\bar{d}^\text{G}/t^\text{GB}$ and the GB conductivity $\sigma^\text{GB}/\sigma^\text{G}$. Both analytical and numerical results are presented. The two RVEs with the smallest and largest grain size are displayed. The grain size distribution within the RVEs are relatively uniform. Note that the image size does not represent the real RVE size. (b) Distributions of electric potential and (c) current density for the four RVEs with different $\bar{d}^\text{G}/t^\text{GB}$ and $\sigma^\text{GB}/\sigma^\text{G}$. The RVEs are subjected to a macroscopic electric field in the $x$-direction. The numerical results agree well with the corresponding analytical results. $\sigma^\star/\sigma^\text{G}$ increases monotonically with $\bar{d}^\text{G}/t^\text{GB}$ and tends to saturate if $\bar{d}^\text{G}/t^\text{GB}$ is sufficiently large. A larger  $\sigma^\text{GB}/\sigma^\text{G}$ yields a larger  $\sigma^\star/\sigma^\text{G}$ and a weaker grain size-dependence. Increasing $\bar{d}^\text{G}/t^\text{GB}$ and/or $\sigma^\text{GB}/\sigma^\text{G}$ results in a smoother electric potential distribution across the GB and a higher level of current density, which attributes for the increase of the effective conductivity. }
	\label{size_effect}
\end{figure}
Due to the GB effect, the effective ionic conductivity of polycrystals is size-dependent. In the supplementary material, the analytical expression of the effective conductivity for an idealized polycrystal system is derived, i.e.,
\begin{equation}
	\frac{\sigma^\star}{\sigma^\text{G}}=\frac{1}{\frac{\sigma^\text{G}}{\sigma^\text{GB}}\frac{t^\text{GB}}{d^\text{G}}+1}.
	\label{an_size_effect}
\end{equation}
The above formula is identical to that from the brick layer model \cite{avila2010protonic,wang2014schottky}. Unlike the brick-layer model where the expression is obtained based on an equivalent circuit \cite{verkerk1982effect}, the expression \eqref{an_size_effect} is derived by analytically solving an boundary value problem consisting of the governing equations for both bulk and GB conduction together with the corresponding boundary conditions. This provides a new way to interpret the brick-layer model. Relevant details are found in the supplementary material.

The analytical solution \eqref{an_size_effect} indicates that the effective conductivity is dominated by the grain size-to-GB thickness ratio, i.e., $d^\text{G}/t^\text{GB}$. Additionally, the GB-to-grain conductivity ratio $\sigma^\text{GB}/\sigma^\text{G}$ also plays a crucial role. Thus, we simulate a number of RVEs with varying normalized nominal average grain size $\bar{d}^\text{G}/t^\text{GB}$ and $\sigma^\text{GB}/\sigma^\text{G}$. The normalized effective conductivity $\sigma^\star/\sigma^\text{G}$ varying with normalized grain size $\bar{d}^\text{G}/t^\text{GB}$ and the GB conductivity $\sigma^\text{GB}/\sigma^\text{G}$ are plotted in Fig.\,\ref{size_effect}a. For comparison, the corresponding results from the analytical solution \eqref{an_size_effect} are also displayed. 

The numerical and analytical results match well with each other, indicating that the simple analytical expression \eqref{an_size_effect} can already well capture the grain size-dependence of the effective ionic conductivity. $\sigma^\star/\sigma^\text{G}$ exhibits a monotonic increase with the average grain size  $\bar{d}^\text{G}/t^\text{GB}$ and approaches saturation as $\bar{d}^\text{G}/t^\text{GB}$ becomes sufficiently large. This tendency is highly consistent with the experimental measurements \cite{wu2017origin}. $\sigma^\star/\sigma^\text{G}$ increases with $\sigma^\text{GB}/\sigma^\text{G}$. Moreover, in the case with a larger $\sigma^\text{GB}/\sigma^\text{G}$, the grain size-dependence of the effective conductivity is weaker. In other words, the grain size effect can be compensated by increasing the GB conductivity. In turn, increasing the grain size can weaken the detrimental effect on the effective conductivity due to the low GB conductivity.

The macroscopic grain size-dependence of the effective ion conductivity is attributed to the underlying microscopic ionic conducting behaviors. As shown in Figs.\,\ref{size_effect}b and \ref{size_effect}c, increasing $\bar{d}^\text{G}/t^\text{GB}$ and/or $\sigma^\text{GB}/\sigma^\text{G}$ results in a smoother electric potential distribution across the GB and a higher level of current density and hence a higher effective ionic conductivity. 

From the above discussion, the effective ionic conductivity can be enhanced by increasing either the average grain size or the GB conductivity. In this regard, the analytical and numerical results above provide quantitative guidelines.

In the above cases with relatively homogeneous distribution of grain and GB features, the simple analytical solution \eqref{an_size_effect} agrees well with the numerical simulations. However, only numerical simulations are capable to tackling cases involving an inhomogeneous distribution of grain and GB features. In fact, similar conclusions have been made in some other works, e.g., \cite{fleig1999impedance,dezanneau2006grain,dezanneau2006effect,fleig2002grain}.
\subsection{Grain orientation-dependence of effective ionic conductivity}
\begin{figure}[!htbp]
	\centerline{\includegraphics[width=\textwidth]{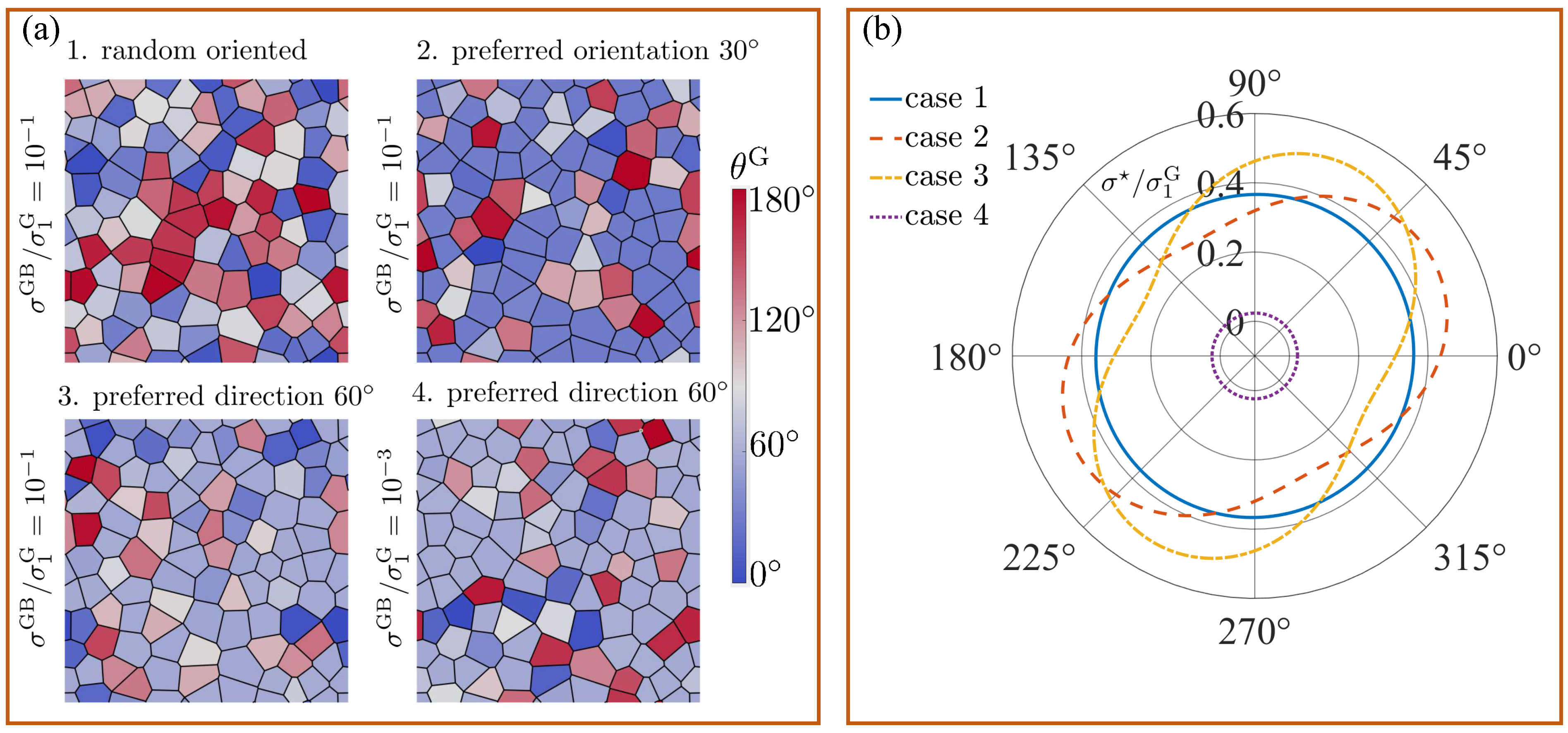}}
	\caption{Grain orientation-dependent effective conductivity. (a) Distributions of grain orientation. (b) Polar plots of the orientation-dependent effective ionic conductivity for the four cases. For case 1, the orientation distribution is random. For the other three cases, a preferred orientation angle $\theta^\text{G}_\text{p}$ is prescribed, which is $30^\circ$ for case 2 and $60^\circ$ for cases 3 and 4. About 50\% of the grains orientate towards the preferred orientation, i.e., possess an orientation angle $\theta^\text{G}_\text{p}\pm 2^\circ$. These three cases represent the polycrystalline microstructures with a texture. Case 4 possesses a much lower GB conductivity than the other three cases. The other relevant parameters are $\sigma^\text{G}_{1}/\sigma^\text{G}_{2}=5$ and $\bar{d}^\text{G}/t^\text{GB}=40$. The effective conductivity is almost isotropic in cases 1 and 4 and is strongly anisotropic with the maximum appearing along the preferred orientation in cases 2 and 3. These results indicate that the orientation-dependence of the effective conductivity relies on the grain orientation distribution as well as the GB conductivity.}
	\label{orientation}
\end{figure}

Here, we investigate how the grain orientation affects the effective ionic conductivity. To this end, as shown in Fig.\,\ref{orientation}, we consider four cases. The grain orientation distributions are plotted in Fig.\,\ref{orientation}a. The orientation distribution in case 1 is random. In the other three cases, to consider the influence of the texture \cite{messing2017texture}, a preferred orientation is prescribed such that the majority of the grains possess a similar orientation. In addition, to investigate the GB effect, a much lower GB conductivity is taken for case 4. The grain bulk conductivity is considered to be anisotropic, i.e., $\sigma^\text{G}_{1}/\sigma^\text{G}_{2}=5$. The polar plots of the effective conductivity are displayed in Fig.\,\ref{orientation}b.
Despite the high degree of anisotropy of the bulk conductivity, the effective conductivity in case 1 is almost isotropic, which is due to the random grain orientation. In cases 2 and 3, the effective conductivity is strongly anisotropic with the maximum value appearing along the preferred orientation. The maximum effective conductivity is much larger than the isotropic one in case 1. The effective conductivity in case 4 is much smaller than the other three cases due to the much lower GB conductivity. But it is almost isotropic, which is explained as follows. As the GB conductivity is much lower than the bulk conductivity, the effective conductivity is governed by the ionic conducting behavior at the GB. Therefore, the random distribution of the GB orientation (see Fig.\,\ref{orientation}a) results in the isotropy of the effective conductivity. In this case, the GB effect compensates the influence of grain orientation. 

According to the above investigations, in the case with an anisotropic bulk conductivity, the degree of anisotropy of the effective conductivity can be tuned by controlling the grain orientation distribution. Particularly, introducing a preferred orientation (i.e., a texture) is beneficial for achieving a high conductivity along a specific direction, e.g., the current flow direction. Thus, the results presented here can guide the engineering of grain orientations in practice \cite{messing2017texture}.

It is worth mentioning that the influence of grain orientations on the effective conductivity in polycrystals have been investigated by analytical methods in some works (e.g., \cite{hashin1963conductivity,lavrov2011effective}). However, these analytical methods ignore the influence of GBs and only applicable to idealized distribution of grain orientations. In contrast, the current numerical framework with consideration of GB conduction is general and capable of analyzing both artificially generated and experimental measured (e.g., \cite{sharafi2017controlling,fu2023grain}) grain orientations.
\subsection{GB conductivity-dependence of effective ionic conductivity}
Depending on the GB characters and the local chemical compositions, the ionic conductivity may differ among GBs in a certain polycrystal \cite{fleig1999impedance,bowman2019linking,mcnealy2014use}. Thus, the distribution of GBs with low and high conductivities is also of great importance. In fact, as suggested by \cite{bowman2019linking}, the percolation of GBs with high conductivities can significantly enhance the effective conductivity. Here, the influence of the GB conductivity distribution is considered. Without loss of generality, we classify the GBs into low-conducting and high-conducting GBs \footnote{In real materials, the distribution of GB conductivity could be more complicated \cite{bowman2019linking}. However, it may still be reasonable to classify GBs into low conducting and high conducting types. The present model and numerical methods are capable of analyzing cases with a general distribution of GB conductivities such as the normal distribution considered in \cite{mcnealy2014use}.}. Their normalized GB conductivities $\sigma^\text{GB}/\sigma^\text{G}$ are $10^{-3}$ and $10^{-1}$, respectively. 
\begin{figure}[!htbp]
	\centerline{\includegraphics[width=\textwidth]{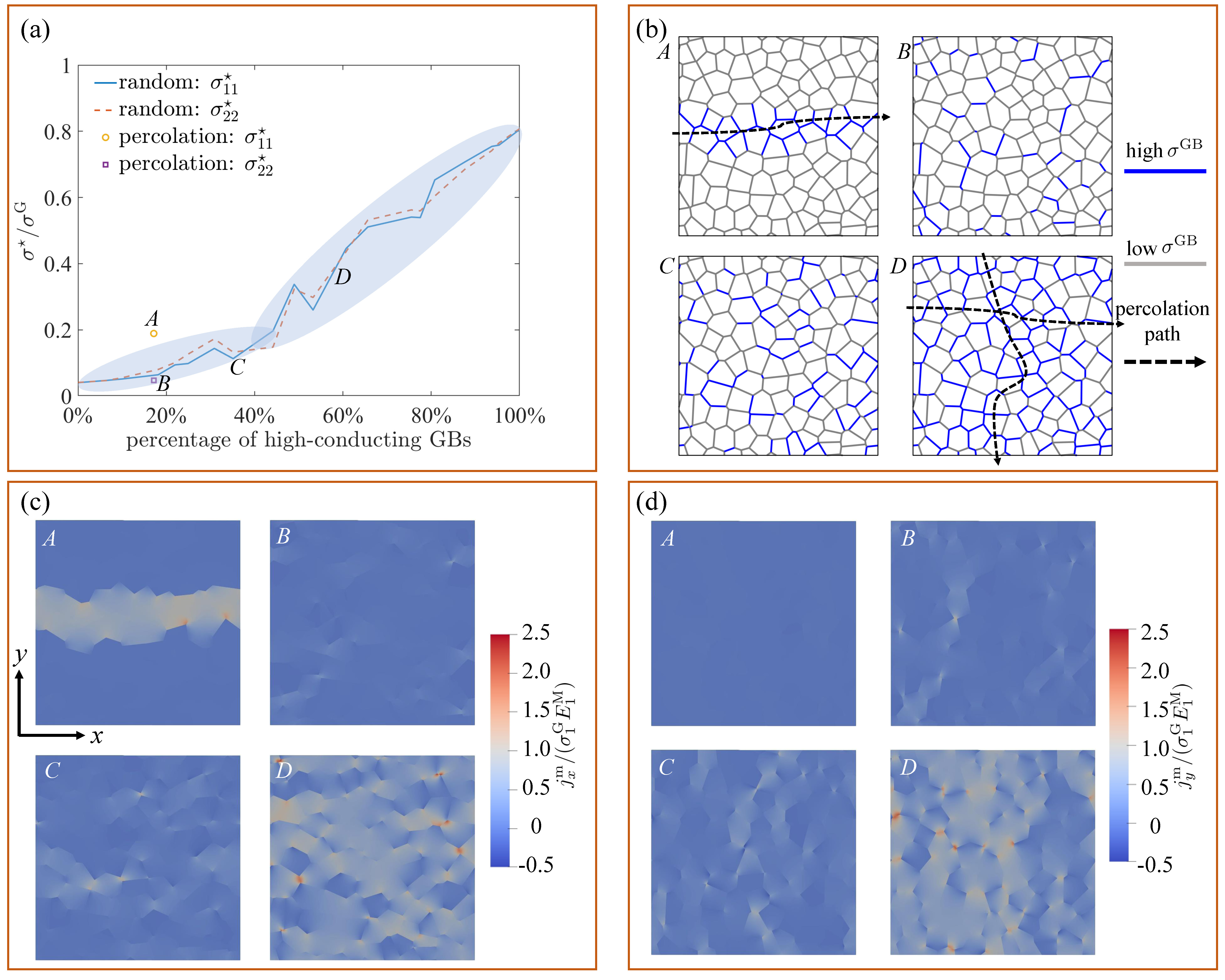}}
	\caption{Influence of GB conductivity on the effective conductivity. (a) The normalized effective conductivities $\sigma^\star_{11}/\sigma^\text{G}$ and $\sigma^\star_{22}/\sigma^\text{G}$ varying with the percentage of high-conducting GBs. The percentage is defined as the total length of the high-conducting GBs divided by the total length of all GBs. For all the cases except the one with a prescribed percolation path (i.e., the two markers), the high-conducting GBs are randomly distributed. (b) Distributions of  high-conducting GBs for the four cases marked in the plots in (a).  (c) and (d) Distribution of current density. The RVEs in (c) and (d) are subjected to a macroscopic electric field in the $x$- and $y$-directions, respectively. The average grain size is $\bar{d}^\text{G}/t^\text{GB}=40$. The bulk conductivity is isotropic, i.e., $\sigma^\text{G}_{1}/\sigma^\text{G}_{2}=1$. The effective conductivities tend to increase with the percentage of high-conducting GBs. Meanwhile, the arrangement of high-conducting GBs is also of great importance. The existence of the percolation can significantly enlarge the effective conductivity. The threshold value for percolation is around 40\%, by which the curves are divided into two ranges (marked by the two transparent ellipses). The second range with percolation exhibits a much higher increasing rate.}
	\label{GB_distribution}
\end{figure}
In Fig.\,\ref{GB_distribution}a, the normalized effective conductivities $\sigma^\star_{11}/\sigma^\text{G}$ and $\sigma^\star_{22}/\sigma^\text{G}$ in the $x$- and $y$-directions varying with the percentage of high-conducting GBs is plotted. For all the cases except the one with a prescribed percolation path (i.e., the two markers), the high-conducting GBs are randomly distributed. The effective conductivities tend to increase with the percentage of high-conducting GBs. The fluctuations on the curves indicate that a smaller percentage but a better arrangement of high-conducting GBs can result in a larger effective conductivity. This phenomenon is more evident when comparing case A with case B. Both cases possess about 18\% high-conducting GBs, while the effective conductivity in the $x$-direction in case A is about 3 times that in case B. This is attributed to the prescribed percolation path along the $x$-direction in case A (see Fig.\,\ref{GB_distribution}b). The effective conductivities in the $x$- and $y$-directions are not always the same. This anisotropic behavior is induced by the uneven distribution of high-conducting GBs such that the conducting behavior is different in the two directions as suggested by the distribution of current density in Figs.\,\ref{GB_distribution}c and \ref{GB_distribution}d. 

There exists a threshold value of the percentage of high-conducting GBs, by which the effective conductivity versus percentage curves in Fig.\,\ref{GB_distribution}a are divided into two ranges with distinctive increasing rates (i.e., the nominal slope). Seen from the curves, the threshold value is around 40\% for the current problem. For cases B and C in the first range with a percentage smaller than the threshold value, the percolation does not occur. For case D in the second range with a percentage larger than the threshold value, the percolation takes place. The existence of percolation magnifies the contribution of high-conducting GBs, which leads to a much higher increasing rate of the effective conductivity in the second range. The influence of the percolation is also reflected by the distribution of current density in Figs.\,\ref{GB_distribution}c and \ref{GB_distribution}d. The current density along the percolation path is much higher, indicating that it is a highway for ionic conducting (see cases A and D).

In a short conclusion, to achieve a higher effective conductivity, it is necessary to have at least a certain percentage of high-conducting GBs. Moreover, to optimize the effective conductivity,  it is crucial to engineer their distribution to form percolation paths. 

\subsection{GNN-based surrogate modeling}
In the above sections, we have systematically investigated the influence of various grain and GB-relevant features on the effective conductivity by FE simulations, which results in important guidelines for optimizing the effective conductivity. Here, we construct and train a GNN-based ML model to encapsulate these important microstructure-property correlations. The details on the GNN model are found in Section~\ref{surrogate model}.
\begin{figure}[!htbp]
	\centerline{\includegraphics[width=\textwidth]{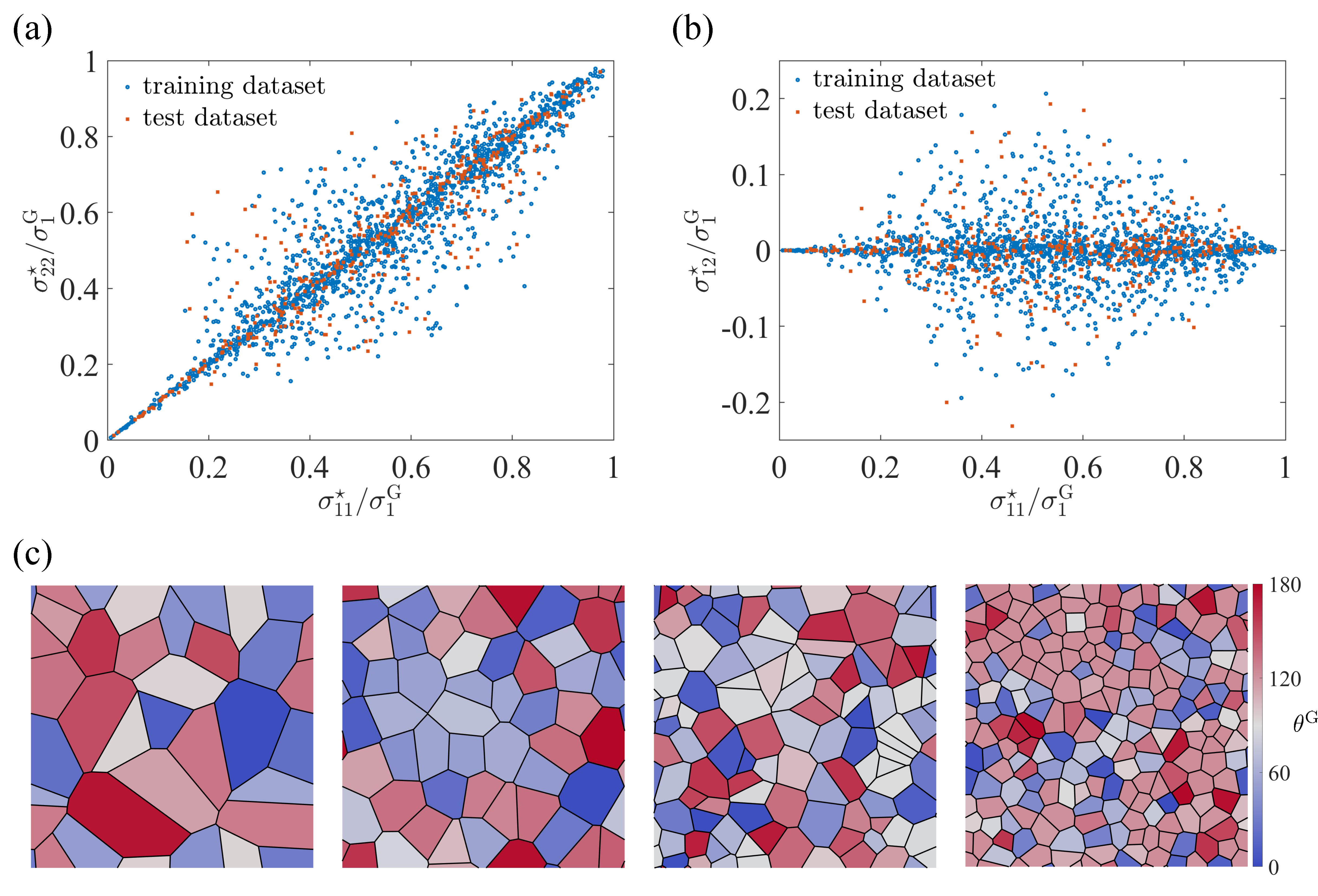}}
	\caption{Training and test datasets. The effective conductivities of polycrystals in the training and test datasets are plotted in the form of two combinations: (a) $\sigma^\star_{11}/\sigma^\text{G}_{1}$ versus $\sigma^\star_{22}/\sigma^\text{G}_{1}$ and (b) $\sigma^\star_{11}/\sigma^\text{G}_{1}$ versus $\sigma^\star_{12}/\sigma^\text{G}_{1}$. (c) The four typical RVEs from the dataset with different grain numbers, degrees of regularity, and grain orientation distribution. We generate 2000 RVEs in total by varying the grain and GB parameters. Their effective conductivities are evaluated by FE simulations. These result in 2000 data points. 80\% of them belong to the training dataset for the model training and the rest 20\% are in the test dataset for the model validation.}
	\label{data_set}
\end{figure}
\begin{figure}[!htbp]
	\centerline{\includegraphics[width=\textwidth]{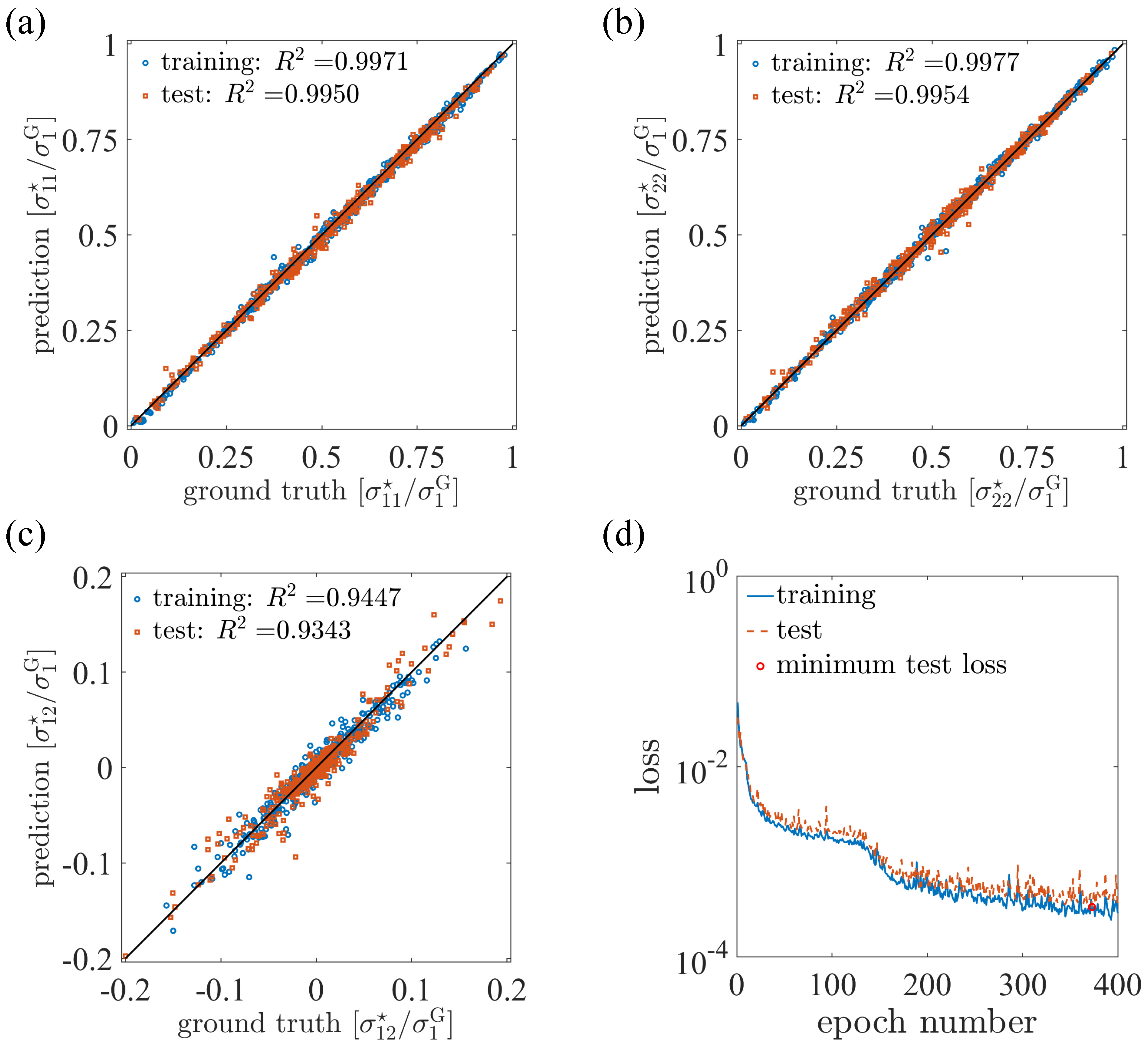}}
	\caption{GNN model performance. The effective conductivities predicted by the trained GNN model in comparison with those from FE simulations:
		(a) $\sigma^\star_{11}/\sigma^\text{G}_{1}$, (b) $\sigma^\star_{22}/\sigma^\text{G}_{1}$, and (c) $\sigma^\star_{12}/\sigma^\text{G}_{1}$. (d) The training and test loss-epoch curves. The GNN model prediction agrees well with the ground truth from FE simulations. For both the training and test datasets, the coefficients of determination (i.e., $R^2$) for all three conductivity components exceed 0.93. This suggests the good performance of the trained GNN model. The test loss-epoch curve matches with the training one, indicating that the GNN model is well-trained without noticeable underfitting or overfitting. The slight oscillating on the curves makes
		the last epoch not necessarily to be the one with the smallest loss. Thus, the GNN model after each epoch is saved and the one with the
		smallest test loss (marked the red circle in (d)) is taken as the converged model.}
	\label{performance}
\end{figure}
The data generation is discussed first. We generate 2000 RVEs by varying the grain and GB features including the number of grains in an RVE (25 to 196), the nominal average grain size $\bar{d}^\text{G}/t^\text{GB}$ (10 to 1000), the distribution of grain orientation (fully random to fully specified), the degree of regularity (almost regular to fully irregular), the GB-to-grain conductivity ratio $\sigma^\text{GB}/\sigma^\text{G}$ ($10^{-4}$ to $10^{-1}$), and the ratio between the two principal bulk conductivities $\sigma^\text{G}_2/\sigma^\text{G}_1$ (0.1 to 1). The effective conductivities of each RVE are evaluated by FE simulations. The resulting dataset consists of 2000 RVEs paired with the corresponding effective conductivities. 80\% of them are used for training the model and the rest 20\% are for the model validation. As they vary greatly in magnitude, both inputs and outputs are scaled to the range $[0,\,1]$ by the min-max scaling. The effective conductivities from both datasets are shown in Figs.\,\ref{data_set}a and \ref{data_set}b. The four exemplary RVEs with the corresponding grain orientation distribution are displayed in Fig.\,\ref{data_set}c. The dataset is representative and hence is suitable for the GNN model training.

The performance of the trained GNN model is illustrated in Fig.\,\ref{performance}. A good agreement between the GNN model prediction and the ground truth from FE simulations is observed. The coefficients of determination (i.e., $R^2$) for the two axial components $\sigma^\star_{11}/\sigma^\text{G}_{1}$  and $\sigma^\star_{22}/\sigma^\text{G}_{1}$ exceed 0.99 in both the training and test cases. That for $\sigma^\star_{12}/\sigma^\text{G}_{1}$ is smaller but still larger than 0.93.  The good match between the test loss-epoch curve and the training one suggests that the GNN model is well-trained without underfitting or overfitting. These results together indicate that the trained GNN model performs well. 

The GNN model can act as an efficient property predictor. It takes the defined grain and GB features of a polycrystalline microstructure as inputs and predicts the corresponding effective ionic conductivity. To determine the effective conductivity of a microstructure, it takes about $10\,\text{s}$ by FE simulations. With the same computing resource, the time required for the GNN model is about $0.0015\,\text{s}$. Thus, the GNN model is much more efficient than FE simulations. This efficient surrogate model can be combined with traditional optimization methods (e.g., \cite{karapiperis2023prediction}) to optimize the performance of ionic conducting polycrystalline materials. The dataset and trained GNN model are found at \url{https://github.com/XiangLongPeng/GNN_polycrystal_ion_conducting_ceramics}.
\section{Conclusion}
\label{Conclusion}
In this work, we investigate the effective ionic conductivity of polycrystalline oxide ceramics. An ionic conducting model with consideration of the GB effect is introduced. Its numerical implementation is realized by the FE method, which is verified by a benchmark problem with an analytical solution. The computational homogenization method is adopted to calculate the effective ionic conductivity by conducting FE simulations at the RVE level. The influence of grain and GB features including the grain size, grain orientation, and GB conductivity on the effective ionic conductivity is systematically studied. These investigations provide the following guidelines on improving the effective conductivity:
\begin{itemize}	
	\item[$\bullet$] Increasing the average grain size and/or the GB conductivity leads to the increase of effective conductivity.
	 The detrimental effect of the low GB conductivity can be compensated by increasing the average grain size, and vice versa. This is well captured by the analytical expression \eqref{an_size_effect}.                      
	\item[$\bullet$] The degree of anisotropy of the effective conductivity is tunable by engineering the grain orientation distribution. Introducing a preferred orientation (or texture) results in a high conductivity along a specific direction, which can act as the current flow direction.
	\item[$\bullet$] A certain percentage of GBs with a high conductivity is necessary for attaining a high effective conductivity. Engineering their distribution to form percolation paths is crucial for achieving optimized effective conductivity. 
\end{itemize}
The above microstructure-property correlations are encapsulated in a GNN-based ML model. This surrogate model efficiently predicts the effective ionic conductivity for a given polycrystalline microstructure. Thus, it is a valuable tool for enhancing the performance of ionic conducting polycrystalline oxide ceramics by microstructure engineering. 

In this work, we only consider artificially generated microstructures. The model and numerical method proposed here is also applicable to real microstructures with experimentally measured features. Here, the forward prediction task, i.e., evaluating the effective properties of a known microstructure is tackled. The inverse design and optimization of microstructures are also of great importance. One possible way to realize microstructure optimization is to combine the ML model with traditional optimization methods (e.g., \cite{karapiperis2023prediction}). These aspects are left for future investigations.

\section*{Data availability}
The data required to reproduce findings of this work are found at \url{https://github.com/XiangLongPeng/GNN_polycrystal_ion_conducting_ceramics}.
\section*{Acknowledgments}
The authors gratefully acknowledge the computing time granted on the Hessian High-Performance Computer ‘‘Lichtenberg”.
\bibliographystyle{elsarticle-num}


\end{document}


\begin{frontmatter}

\title{Supplementary Material: \\Enhancing ionic conductivity of ceramic electrolytes by microstructure engineering: computational homogenization and machine learning}
\author[tud]{Xiang-Long Peng\corref{cor1}}
\ead{xianglong.peng@tu-darmstadt.de}
\author[tud]{Bai-Xiang Xu}
\cortext[cor1]{Corresponding author}
\address[tud]{Mechanics of Functional Materials Division, Institute of Materials Science, Technische Universität Darmstadt, 64287 Darmstadt, Germany}

\end{frontmatter}

\renewcommand{\thefigure}{S.\arabic{figure}}
\renewcommand{\thetable}{S.\arabic{table}}
\renewcommand{\theequation}{S.\arabic{equation}}

\section{A bicrystal benchmark problem: analytical versus numerical solutions}
\begin{figure}[!htbp]
	\centerline{\includegraphics[width=\textwidth]{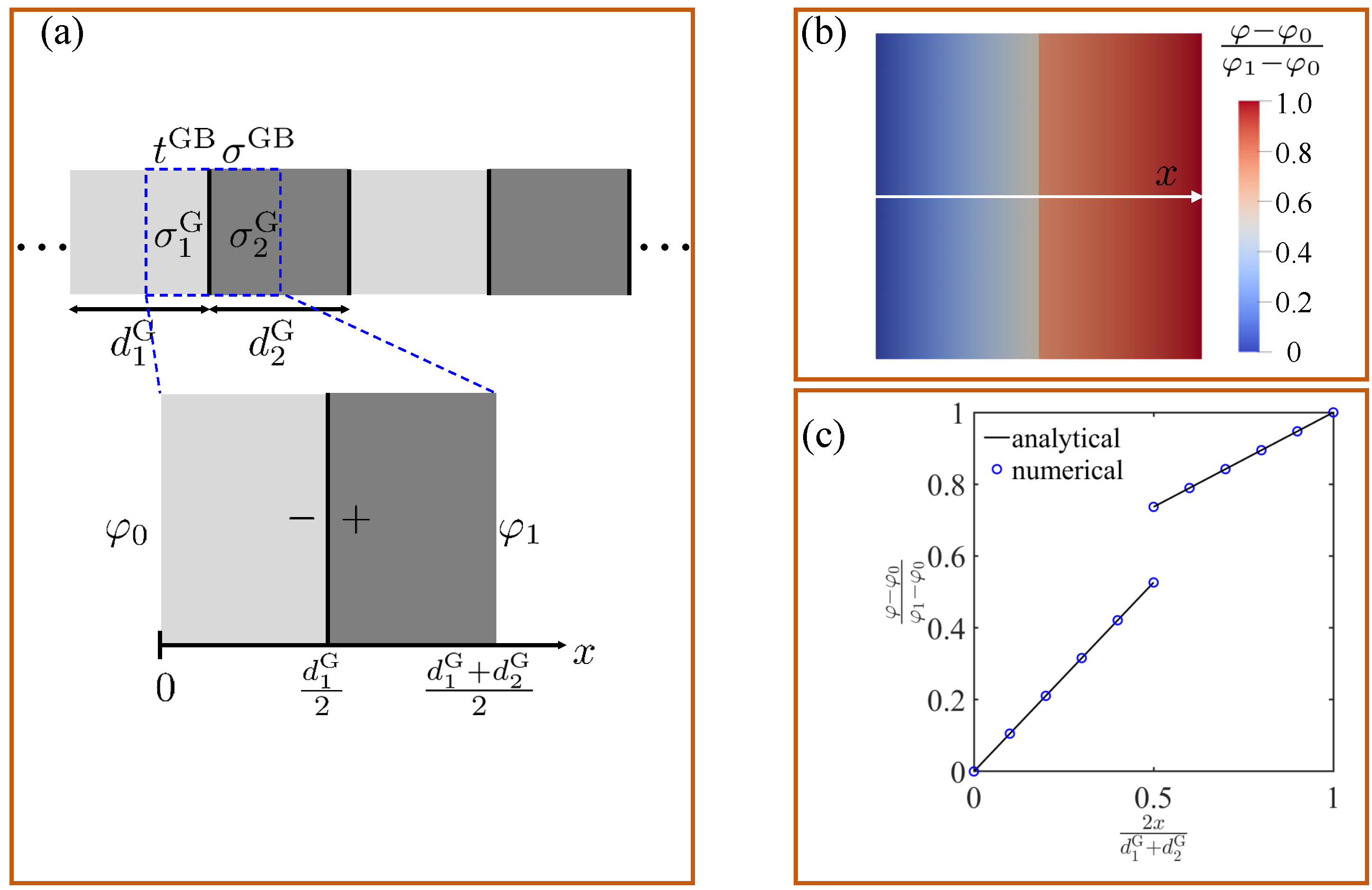}}
	\caption{A bicrystal benchmark problem. (a) The schematic of the periodic bicrystal system and one-half of the unit cell of interest. (b) The contour plot of the electric potential distribution from the FE simulation. (c) The distribution of electric potential in the $x$-direction from numerical simulation in comparison with the analytical solution. The distribution of the electric potential is homogeneous in the $y$-direction and is linear within each grain in the $x$-direction. Across the GB, the electric potential is subjected to a jump. The perfect agreement between the numerical result and analytical solution verifies the numerical implementation. The relevant parameters are $\sigma^\text{G}_1/\sigma^\text{GB}=20$, $\sigma^\text{G}_2/\sigma^\text{GB}=40$ and $d^\text{G}_1/t^\text{GB}=d^\text{G}_2/t^\text{GB}=100$.}
	\label{analy_numer}
\end{figure}
We adopt a benchmark problem to verify the numerical implementation of the model. As illustrated in Fig.\, \ref{analy_numer}a, we consider a layered polycrystal system consisting of periodically repeated bicrystal unit cells. The ionic conducting behavior in the periodic direction (i.e., $x$-direction) is of concern. For such a one-dimensional problem, it is sufficient to consider one-half of the unit cell. The grain sizes of the two grains in the bicrystal are $d^\text{G}_1$ and $d^\text{G}_2$, respectively. Their ionic conductivities in the $x$-direction are $\sigma^\text{G}_{1}$ and $\sigma^\text{G}_{2}$. The conductivity and thickness of the GB between them are $\sigma^\text{GB}$ and $t^\text{GB}$, respectively. The bicrystal is subjected to the following Dirichlet boundary conditions at both ends, i.e., 
\begin{equation}
	\varphi(0)=\varphi_0,\quad \varphi(\frac{d^\text{G}_1+d^\text{G}_2}{2})=\varphi_1.
\end{equation}
In the bulk, we have the following governing equations
\begin{equation}
	-\sigma^\text{G}_{1}\frac{\partial\varphi}{\partial x}=0,\quad 0\leq x\leq\frac{d^\text{G}_1}{2}
\end{equation}
and
\begin{equation}
	-\sigma^\text{G}_{2}\frac{\partial\varphi}{\partial x}=0,\quad \frac{d^\text{G}_1}{2}\leq x\leq\frac{d^\text{G}_1+d^\text{G}_2}{2}.
\end{equation}
At the GB, we have
\begin{equation}
	-\sigma^\text{G}_{1}\left[\frac{\partial\varphi}{\partial x}\right]^{-}=-\frac{\sigma^\text{GB}}{t^\text{GB}}\left[\varphi^{+}(\frac{d^\text{G}_1}{2})-\varphi^{-}(\frac{d^\text{G}_1}{2})\right],
\end{equation}
and
\begin{equation}
	-\sigma^\text{G}_{2}\left[\frac{\partial\varphi}{\partial x}\right]^{+}=-\frac{\sigma^\text{GB}}{t^\text{GB}}\left[\varphi^{+}(\frac{d^\text{G}_1}{2})-\varphi^{-}(\frac{d^\text{G}_1}{2})\right],
\end{equation}

The analytical solution to the above boundary-value problem can be readily obtained. The analytical expression for the distribution of electric potential $\varphi(x)$ is
\begin{equation}
	\varphi(x)=
	\left\{ \begin{array}{ll}
		\frac{C_1(\varphi_1-\varphi_0)}{t^\text{GB}+\frac{C_{1}d^\text{G}_1}{2}+\frac{C_{2}d^\text{G}_2}{2}}x+\varphi_0, & 0\leq x\leq\frac{d^\text{G}_1}{2} \\
		\\
		\frac{C_2(\varphi_1-\varphi_0)}{t^\text{GB}+\frac{C_{1}d^\text{G}_1}{2}+\frac{C_{2}d^\text{G}_2}{2}}\left(x-\frac{d^\text{G}_1+d^\text{G}_2}{2}\right)+\varphi_1,
		&\frac{d^\text{G}_1}{2}\leq x\leq\frac{d^\text{G}_1+d^\text{G}_2}{2}
	\end{array} \right. 
\end{equation}
with $C_1=\sigma^\text{GB}/\sigma^\text{G}_1$ and $C_2=\sigma^\text{GB}/\sigma^\text{G}_2$. With the aid of the distribution of electric potential,
one can calculate the effective electric field $E^\star$ and effective current density $j^\star$ in the $x$-direction, i.e.,
\begin{equation}
	E^\star=-\frac{2(\varphi_1-\varphi_0)}{d^\text{G}_1+d^\text{G}_2},\quad j^\star=-\frac{\sigma^\text{GB}(\varphi_1-\varphi_0)}{t^\text{GB}+\frac{C_{1}d^\text{G}_1}{2}+\frac{C_{2}d^\text{G}_2}{2}}
\end{equation}
Then,  the effective ionic conductivity is obtained, i.e.,
\begin{equation}
	\sigma^\star=\frac{j^\star}{E^\star}=\frac{\sigma^\text{GB}(d^\text{G}_1+d^\text{G}_2)}{2t^\text{GB}+C_{1}d^\text{G}_1+C_{2}d^\text{G}_2}.
\end{equation}
By assuming that both grains possess the same parameters, i.e., $\sigma^\text{G}_1=\sigma^\text{G}_2=\sigma^\text{G}$ and $d^\text{G}_1=d^\text{G}_2=d^\text{G}$, the above equation is rewritten as
\begin{equation}
	\frac{\sigma^\star}{\sigma^\text{G}}=\frac{1}{\frac{\sigma^\text{G}}{\sigma^\text{GB}}\frac{t^\text{GB}}{d^\text{G}}+1}.
\end{equation}
From the above analytical solution, the normalized effective conductivity $\sigma^\star/\sigma^\text{G}$ depends on the GB-to-grain conductivity ratio $\sigma^\text{GB}/\sigma^\text{G}$ and the grain size-to-GB thickness ratio $d^\text{G}/t^\text{GB}$.

The above benchmark problem is also solved numerically by the FE method described in the main text. As shown in Fig.\,\ref{analy_numer}c. The distribution of the electric potential from the numerical simulation perfectly agrees with the analytical one, which verifies the numerical implementation of the model.